# Towards a Contemporary Definition of Cybersecurity

Francesco Schiliro' [1, 2]

f.schiliro@adfa.edu.au

[1]Australian Defence Force Academy, University of New South Wales, Canberra
[2]Macquarie University

21 November 2022

**Abstract**

The report provides an intricate analysis of cyber security defined in contemporary operational digital environments. An extensive literature review is formed to determine how the construct is reviewed in modern scholarly contexts. The article seeks to offer a comprehensive definition of the term "cybersecurity" to accentuate its multidisciplinary perspectives. A meaningful concise, and inclusive dimension will be provided to assist in designing scholarly discourse on the subject. The report will offer a unified framework for examining activities that constitute the concept resulting in a new definition; "Cybersecurity is the collection and concerting of resources including personnel and infrastructure, structures, and processes to protect networks and cyber- enabled computer systems from events that compromise the integrity and interfere with property rights, resulting in some extent of loss." The encapsulation of the interdisciplinary domains will be critical in improving understanding and response to emerging challenges in the cyberspace.

**Key terms:** Cybersecurity, prevention detection, incident, response, interdisciplinary, malicious actor, computer systems, infrastructure, protection, interference, property rights

**Defining Cybersecurity/Cybercrime**

The term "cybersecurity" is commonly used to refer to a set of circumstances or events related to improving the integrity of a given information management system or infrastructure and addressing present and emerging challenges associated with the exercise. The concept is often broadly invoked to refer to any steps taken to ensure the security of information technology resources. The regulation is often inadequate as it fails to capture the entire expanse of the fore and background activities commonly involved in the adventures. A comprehensive and intimate analysis of the term is not often provided in existing literature, further perpetuating the misapplication of the concept in contemporary information management settings. A more unified definition should entail the collection and organization of resources, personnel, and processes to prevent, detect, isolate and resolve potential threats to the continued well-being of information technology infrastructure.

The contemporary digital environment is increasingly integrating the internet of things to improve connectivity and interaction between different sectors of people's lives. The concept has become commonplace due to its inherent benefits, including error reduction and improvement of efficiency by minimizing continued human involvement. Increased connectivity has effectively eliminated traditional challenges associated with manual input, thus, exhibiting increased performance. The expansive integration of information technology in virtually every aspect of existence has greatly impacted conventional processes, including increased efficiency and enhanced productivity. Concepts such as data analytics and other information management strategies have been consequential in defining policy discourse and expanding understanding of potential quality improvement areas in every industry. However, the increased interconnectivity continues to present a major risk to the well-being of the data processing and management superstructure on which industries have grown exceedingly reliant. Cybersecurity is a major concern in the information technology (IT) environment and is one of the most important aspects of implementing any IT infrastructure. The concept involves identifying potential vulnerabilities and instituting effective strategies to minimize the effect of possible threats. Cybersecurity is a quality improvement initiative adopted by organizations to monitor and ensure the integrity of their systems against unauthorized entry (Wong, 2007).

However, while the term is often invoked in many information technology and data management settings including by its explicit definition and underpinning principles are rarely discussed. The report will provide an in-depth examination of the concept, including exploring its origin and historical application and reviewing its evolution over the past decades and adoption during the digital era. The construct's implications in the contemporary information technology environment will also be explicitly discussed, along with its underpinning best practices.

**Literature Review**

**Prevention, Detection, and Response**

Cybersecurity has been extensively examined in existing literature, with the different facets of the concept being reconstructed and analysed in light of the evolving contextual factors in the contemporary operational environment. A review of the literature by the National Institute of Standards and Technology (NIST) reveals that construct can be analysed by categorizing it into its constituent elements, including the internet of things, cloud, network, application, and critical infrastructure security (Cybersecurity, Critical Infrastructure, 2018). The internet of things is an emerging practice involving the interconnectivity of different systems to optimize their operations and management. NIST posits that the concept presents a significant integrity challenge due to malicious actors' numerous potential points of access and the huge adverse impact of such activity (Cybersecurity, Critical Infrastructure, 2018). The definition of cybersecurity is also increasingly involved in cloud capabilities due to the evolving user behaviour and the expanded preference for offshore storage and computation competencies.

The benefits of cloud computing and its integration into contemporary practice have increased the preference for off-site systems. Beqiri (Beqiri, 2018) explain that networks are a critical aspect of any computer infrastructure and form the basis of its operability. The author supposes that securing a given architecture often requires an advanced understanding of each constituent system, its users, potential vulnerabilities, and other elements that may impact its sustained use. Application is another core categorization of cybersecurity that entails scrutinizing the software functions and their potential to be exploited by malicious actors or impacted by irresponsible use on the client end. The programs may be in-house or vendor-supplied and may present unique challenges in ensuring the integrity and sustaining of organizational operations, especially if compromised. According to Maglaras, et al. (Maglaras, et al., 2018), an analysis of cybersecurity also envelopes reviewing critical infrastructure. The physical integrity of the premises is one of the foremost considerations before implementing any computer system.

Maglaras et al. (Maglaras, et al., 2018) assert that critical infrastructure must often be secured to eliminate the prospect of unauthorized intrusion. A review of the literature also provides in-depth insight regarding the core elements of information security, disaster recovery, planning, network, and end-user (Srinivasan & Simna, 2017). The multidisciplinary nature of cybersecurity analysis includes a review of its major constituent aspect, including end-user activity and other machine-human interactions that improve its discernment. The derivation of a more unified definition of the concept is predicated on a discussion of the dominant themes distinguishing its core aspects. Enriched focus is placed on deconstructing cybersecurity as a multifaceted construct rather than an abstract concept. The term is analysed holistically using a multidisciplinary outlook. A dialectical approach must always be assumed in reviewing cybersecurity as it applies in the contemporary operational environment.

**Theoretical Framework**

An analysis of cybersecurity requires an intimate understanding of the key factors underpinning the practice. Exploring the concept often requires an intimate analysis of the core contextual factors that define its application within the given environment. There has been a sustained attempt by scholars and experts to develop a unified understanding of cybersecurity due to the wide diversity of context-specific nature of the pertinent issues. Several theories have been developed to improve understanding of the concept and to define scholarly discussions regarding cybersecurity amongst government, law enforcement, and industry. The frameworks have been integral to analysing the different facets underpinning the practice. Many of the concepts are often centred on discussing vulnerabilities and motivations of malicious actors, which are largely responsible for the sustained expression of information technology insecurity. The concepts include social cognitive theory, grounded theory, general deterrence theory, distributed cognition theory, activity theory, and social-technical systems theory. The frameworks have been integral to demystifying the various facets that contribute to the implications of cybersecurity in the conventional business environment.

Tamrin et al. (Tamrin, Norman, & Hamid, 2021) explain that the social cognitive theory posits that users' buyer inclination to willingly share information is determined by a plethora of factors, including their expectations, beliefs, and subjective norms. The widespread adoption of social media and the liberalization of personal data sharing has amplified cybersecurity concerns due to the laissez-faire attitude of individuals towards potential concerns presented by their digital conduct. The distribution cognition theory is also instrumental in understanding the concept as it defines how information is transferred and disbursed amongst individuals in a given social setting. The concept is integral to redefining the evolution of best practices and the tendency of people to react to emerging threats. Social organization dictates that define people's perception towards the construct of data security are also explicitly analysed.

Data security is a core aspect of the distribution cognition theory detailing the increased utilization of artificial intelligence technologies to pattern human thought processes, thus, preventing, predicting, detecting, reporting, and resolving potential threats two digital systems and information technology infrastructure. Existing literature has also explicitly analysed sociotechnical systems theory which posits that the performance and design of a given organizational framework can only be optimized if the human and machine aspects are adeptly integrated and perceived as codependently. The construct is integral to understanding the definition of cybersecurity in the contemporary environment as it discusses emerging best practices in the field and the continuous adoption of information technology as a best practice. The theory maintains that an intricate interaction must occur between the operators and the system, with the mannerisms of each being considered in the implementation and subsequent quality improvement practices. Understanding the concept of cybersecurity is contingent on deconstructing it by analysing the contextual factors underpinning its implementation. The core elements of the construct include the internet of things (IoT), cloud, network, application, and critical infrastructure (Beqiri, 2018). IGI Global defines the term "cyber" as a construct explaining the culture of computers, electronic communication, and virtual networks interacting to accomplish a given digital task (What is Cybersecurity?, n.d.). The concept has been expanded to integrate activities occurring within cyberspace or involving internet use and associated hardware and software. Interactions through computer frameworks and other activities involving the manipulation of information technology infrastructure have also been utilized to define the construct. The term "security" in literature ensures the integrity of a given tangible and intangible system. The concept entails investment in infrastructure and the concerting of resources to isolate and diffuse potential threats that may interfere with the conventional operations of the framework. A review of the literature has resulted in the isolation of several potential definitions for cybersecurity.

1. " The process of securing data networks, electronic systems, mobile devices, servers, and computers from malicious actors" Kaspersky Lab (What is Cyber Security?, n.d.).

2. " The techniques defined by published materials defend and protect a given cyber environment of an organization or a user from a potential attacker" (Seemma, Nandhini, & Sowmiya, 2018).

3. "Cybersecurity is the process of securing programs, networks and systems against digital attacks. The exercise prevents malicious activity aimed at destroying charging while accessing sensitive data, misusing user information, or interrupting everyday operations" Cisco (What is Cybersecurity?, n.d.).

4. "Cybersecurity is a practice of preventing possible threats or malicious attacks seeking to unlawfully access data, damage information or disrupt digital processes. The concept is defined by the type of activities instituted by the attacker" (7 Types of Cyber Security Threats, n.d.)

5. "Cybersecurity is the process of defending data networks, electronic systems, mobile devices, servers, and computers from malicious attacks" Gartner (What is Cybersecurity?, n.d.).

6. "Cybersecurity is the protection of networks and other internet-enabled and connected systems, including software, and hardware, from cyber threats. Organizations and individuals engage in the practice to minimize the impact of potential attackers in disrupting normal operations" (Shea, Gills, & Clark, n.d.).

7. "Cybersecurity is the process of deploying technologies, best practices, policies, and people to protect sensitive infrastructure and critical systems of individuals and organizations" (Reddy & Reddy, 2014).

8. "Cybersecurity is a collection of best practices, including infrastructural changes critical in ensuring the integrity of a computer system. The concept entails protecting critical installations and sensitive information from potential digital attacks. The measures are developed to combat threats from malicious attackers targeting the networks and applications originating from without or within the organization" (Abomhara & Koien, 2015).

9. "Cybersecurity is the organization and collection of resources, processes, and structures used to protect the cyberspace and cyber-space enabled systems from occurrences that misalign de jure from de facto property rights" (Craigen, Diakub-Thibault, & Purse, 2014).

Data security is a core aspect of organizational well-being and continuity, as minimal interruptions to its operations from unauthorized entry are averted. Cyber breaches cost an average of $3.8 million annually, including expenses associated with detecting, responding to, and resolving the events (IBM: Consumers Pay the Price as Data Breach Costs Reach All-Time High, 2022). Losses from organizational reputation damage are also often significant n additional to legal ramifications of the attacks, such as lawsuit settlements (Legal consequences of a cyber attack, 2022). Regulatory fines and other punishments may also be imposed, further diminishing the organization's financial prospects. The criminals commonly target unique user information to perpetrate identity theft or sell the data in underground marketplaces. An understanding of the ramifications of digital attacks is critical in developing an intimate explanation of the concept of cybersecurity. The implications of human behaviour and machine tendencies and how they are exploited are central to defining the construct and examining its constituent elements.

**Prevent**

Prevention is one of the core responsibilities of cyber security. The process entails instituting measures to discourage potential attackers from accessing the system. Investment in specialized infrastructure, including personnel, software, and hardware, is often necessary to achieve this outcome. Protection can be accomplished by implementing physical barriers and controlling network or computer system access. Placing some form of geographical deterrent often results in improved security outcomes due to the reduced potential of unauthorized entry. Improving hardware and software capabilities also has a major impact on addressing existing vulnerabilities and mitigating the effect of potential attacks. Digital facilities such as firewalls and antiviruses examine potential vulnerabilities that can be exploited by malicious actors and resolve them.

**Detect**

Detecting is a critical element of cybersecurity, involving examining for and identifying potential threats. Isolating a threat is integral to terminating the extent of its actions on the system and instituting appropriate measures to mitigate its effects. Detection is an integral aspect of incident response that entails highlighting the organization of a potential and authorized entry or use. The process also protects sensitive infrastructure and isolates part of the entity's operations to prevent a more widespread intrusion. The entire securities ecosystem is reviewed for potential malicious activity that may compromise the network.

**Respond**

Incident response is an essential element of cybersecurity that involves invoking substantive steps to mitigate the effects of malicious activity in a computer system or network. Configuration is a response process that defines specific activities to be instituted by an organization innovation in the event of a suspected intrusion. Responding to threats also involves analysis of the characteristics or behaviours of the malicious actor. Substantive approaches are often assumed to contain and resolve the attack. Key activities may include isolating other parts of the system and deploying countermeasures to diffuse the threat.

During the process of response, it is also important that data is identified and collected for investigation by law enforcement officials and be presented in a manner that will be accepted to support a prosecution in court.

**Major Cybersecurity Domains**

Implementing a cybersecurity quality improvement initiative always entails integrating several layers of protection to minimize the potential impacts of malicious attacks. An analysis of the concept involves understanding basic attacker behaviour, including accessing, changing or destroying data. Examining cybersecurity must entail discussing why and how accessing the system or user data may have an impact and why it is often necessary to prevent it.

Defining cybersecurity requires an intimate analysis of primary domains, all major themes constituting the concept. There is an overarching understanding that the construct comprises five main aspects, including the internet of things, cloud, network, application, and critical infrastructure security (Liu, et al., 2019). The underpinning elements of cybersecurity are critical to developing a wholesome comprehension of the concept and creating a new definition acceptable among scholars and digital practitioners. The themes not only improve understanding of the interdisciplinary nature of cyber integrity but are also essential to providing context critical to the definition process.

**Critical Infrastructure Security**

Critical infrastructure security is the fundamental cybersecurity practice. The concept involves establishing ethical and behavioural codes to prevent and resolve potential vulnerabilities in a given computer system (Abouzakhar, 2013). Critical infrastructure security involves instituting measures that minimize unauthorized access to networks by adopting a multi-layered approach in implementing the framework. The concept involves applying physical security measures such as a perimeter mall, identity verification, access control, and other exercises that may impede or detract potential attackers (Cobb, 2021). Isolation of sensitive networks is also commonly invoked as a countermeasure to potential cybersecurity incidences.

**Application Security**

 Application security is an intricate element of cybersecurity as it involves developing and continuously testing features intended to improve the integrity of software and eliminate vulnerability against known and unknown threats (Yadav, Gupta, Singh, Kumar, & Sharma, 2018). The construct entails examining applications for potential weaknesses that can be exploited by malicious actors and creating substantive patches that prevent unauthorized entry. The concept involves integrating codes that identify potential attacks and institute pre-set resolution measures. Application security details the precise activities to be accomplished by the software and defines the limits of their deployment to prevent changes or repurposing by external parties (Mohammed, Alkhathami, Alsuwat, & Alsuwat, 2021). The process entails regular testing to ensure that the protocols function as intended and that no malicious embedded codes endanger the integrity of the system.

**Network Security**

Network security is a core domain of cybersecurity as it involves eliminating systems and abilities that potential attackers can exploit to disrupt organizational operations (What is Network Security?, n.d.). Protecting the integrity and usability of data during transmission involves instituting processes that prevent, detect, and monitor unauthorized access (Cybersecurity vs Network Security vs Information Security, 2022). Malicious attackers are stopped from entering the system or misusing and modifying information exchanged within it. The construct is integral to understanding the concept of cybersecurity as it explores vulnerabilities present in the transmission of data between terminals.

**Cloud Security**

Cloud security is an essential aspect of cybersecurity due to its increased adoption in the modern digital environment. Cloud computing is increasingly becoming integrated into conventional information management frameworks. Offshore storage capabilities provide immense benefits, including minimal spatial or technical requirements by the user. The process involves using third-party services available over the internet for critical information or managing key digital services and facilities such as databases, servers, and storage (Frankenfield, Mansa, & Schmitt, 2022). Vendors also availed off-site intelligence, analytics, and networking, significantly increasing flexibility and fostering innovation. The additional access terminals present potential vulnerabilities that malicious attackers can exploit.

**The Internet of Things Security (IoT)**

Understanding IoT is integral to redefining cybersecurity due to its expanding use and integration into contemporary organizational operations. The internet of things is an emerging construct involving the interconnection of different devices and networks to increase efficiency by eliminating or minimizing the number of processes involved in accomplishing given tasks (Gillis, n.d.). The concept entails facilitating communication amongst systems by integrating cloud competency and remote management capabilities. The devices can interact among themselves and share information based on input provided by the user. The concept is increasingly becoming commonplace in the contemporary operational environment. Companies are actively exploring integrating the internet of things into everyday activities such as transportation, wellness, and home security. However, the interconnected nature of the IoT presents a major integrity challenge making the integration of cybersecurity concepts necessary (Gillis, n.d.). The practice presents an ideal opportunity for malicious attackers to collect uniquely identifying information on users, including the habits, physical location, interests, and identity from the different terminals connected to the systems. Therefore, exploring the internet of things is integral to developing a basic understanding of cybersecurity and its constituent practices. Elements such as awareness and digital habits essential in securing systems are explicitly defined.

**Defining Elements of Cybersecurity/Prevent, Detect, Respond**

The multifaceted nature of cybersecurity warrants an intricate examination of the key themes critical in its definition. Several distinguishing aspects have been developed to provide a more profound comprehension of the construct. A multidisciplinary approach should be assumed to explore the principal elements of cybersecurity as applied in contemporary practice environments, including:
      1. High level of interconnectedness, change, and extent of interaction
      2. Interdisciplinary aspect, including human-machine interaction
      3. Network actor capability similarities.

A literature review indicated that there existed consensus regarding the interdisciplinary nature of computer systems and the efforts required to secure them. Interaction between different elements of cybersecurity, including the human and machine perspective, is necessary for a more comprehensive analysis to apply (Craigen, Diakub-Thibault, & Purse, 2014). Cyber-attack scenarios are often similar in their execution and intention and entail exploiting vulnerabilities presented by the networks and utilizing facilities available such as connectivity, to accomplish a given malicious agenda. Existing definitions of cybersecurity exhibit a technical bias in that they are predominantly focused on discussing the shortcomings of human-machine interaction that not only enable but encourage their attacks. According to Craigen et al. (Craigen, Diakub-Thibault, & Purse, 2014), minimal effort is placed into understanding the overarching motivations of the attackers. A blanket assumption is made that monetary gains or the desire to disrupt systems for political reasons often underpin the incidences. A broader discussion on the subject is necessary amongst practitioners in scholarly settings and among government, law enforcement, and industry for an improved understanding and evidence-based approach to be adopted in the study of cyber integrity (Craigen, Diakub-Thibault, & Purse, 2014). Contemporary definitions can often be critiqued or faulted by analysing the underpinning variables, including context, human behaviour, response capabilities, and other factors that skew an object of policies of the causative factors of cybersecurity.

**Developing a New Definition**

A review of literature reviews varying definitions of the concept of cybersecurity. The reality precipitates the situation that the application of the concept is contextual, with environmental factors playing a major role in the varied understanding of the construct. The definition of cybersecurity is often subjective and dependent on the dominant themes within a given setting. Consequently, a more unified understanding is necessary to further scholarly discourse and encourage expanded scope of practice. Current definitions and curricula of cybersecurity are centred on intrusion detection and resolution as opposed to understanding the actor's intentions, which is often more pertinent (Education sector is leaving Australians vulnerable to cyber attacks, n.d.). Determining the motivations of the hacker is essential to securing the infrastructure. A multidisciplinary approach is necessary to deconstruct and promote expanded discussion and exploration of cybersecurity.

**Contemporary Definitions of Cybersecurity**

A review of existing literature developed from multidisciplinary sources indicates a significantly varied outlook on the concept of cybersecurity. The modern definition of the construct is predominantly based on professional competencies, employee skills, environmental factors, and other key elements unique to the professionals' circumstances (Petersen, Santso, Wetzel, Smith, & Witte, 2020). The table below summarizes the authors' perspectives based on their professional understanding and competency inclinations. The weaknesses of the definitions are also provided.

| Definition | Weakness/Critique |
|---|---|
| **Cybersecurity involves protecting assets, services, data, and networks from unauthorized access preventing potential compromise, loss, or corruption.** | A limited socio-technical perspective is provided. A basic understanding of the construct of cybersecurity is presented, including an elementary discussion of multidisciplinary facets of system integrity. The motivations of potential attacks are not explicitly defined. A restrictive listing of the elements being protected is provided. |
| **Cybersecurity involves the interaction of several processes intended to improve overall system integrity and protect cyberspace by preventing unauthorized entry and preserving the property rights of the users** | The definition highlighted the underpinning of intentional cybersecurity as applied in most contexts. However, minimal attention was placed on discussing the multidisciplinary nature of the concept. |
| **Cybersecurity constitutes a collection of practices intended to improve a given cyber environment's security, safety, and integrity. Practitioners acquire the necessary prerequisite infrastructure and install software to control access and minimize the prospect of an authorized entry.** | The definition appreciates the interdisciplinary nature of cybersecurity and acknowledges multiple parties' efforts to ensure the integrity of the cyber environment. However, the introduction of a construct of safety further dilutes the underpinned argument. |
| **Cybersecurity is a discipline dedicated to studying and analyzing threats to computer systems and networks and strategies for mitigating them. The domain also discusses the protection of digital assets download the introduction of authorization to preserve users' property rights and safeguard the infrastructure's integrity.** | The definition is apt but inadequate in providing a comprehensive analysis of the interdisciplinary nature of cybersecurity. The introduction of the concept of property rights shifted the discussion to human actors and expanded on the potential motivations of hackers. However, the machine perspective was not discussed, with the definition of inferring that only malicious persons are involved in the practice. However, presenting cybersecurity as a field of study, not an isolated abstract concept, is significant in promoting further educational discourse. |
| **Cybersecurity is defined as the state of ensuring authority over computer** | The concepts of control and authority and explicitly defined, providing a more |
| **systems and information management is retained by permitted parties made by malicious actors.** | profound analysis of the underpinning principles of cybersecurity. The characterization of cybersecurity as a state was its only drawback. |

Table 1: summarizes the authors' perspectives based on their professional understanding and competency inclinations.

**New Definition**

The functional model presented by Craigen et al. (Craigen, Diakub-Thibault, & Purse, 2014) suffices and ought to be adopted as a standard definition of cybersecurity. However, it fails to clearly articulate the extent of property rights loss that may be incurred by the victim. Craigen et al. (Craigen, Diakub-Thibault, & Purse, 2014) provide a profound analysis of the concept. However, it can be further improved to encapsulate the major elements defining cybercrime and related digital and psychosocial behaviour. After a comprehensive analysis of existing literature, I propose the following definition, which includes all the core aspects of cybersecurity, including the commonly overlooked multidisciplinary approach.

> *"Cybersecurity is the collection and concerting of resources including personnel and infrastructure, structures, and processes to protect networks and cyber-enabled computer systems from events that compromise the integrity and interfere with property rights, resulting in some extent of loss."*

The definition is examined as follows:
> ... The collection and concerting of resources, including personnel and infrastructure structures and processes...

The aspect explores the numerous interrelated perspectives and dimensions defining the complex nature of cybersecurity and its core elements. The section also acknowledges the socio-technical interactions that commonly occur between human beings and the systems and amongst the networks, which present potential vulnerability and integrity concerns. An intentional effort is placed towards obscuring the exact resources, infrastructure, or processes in appreciation of the dynamic nature of cybersecurity.

> .... To protect networks in cyber-enabled computer systems...

This domain discusses the protection of the systems against all types of threats from within and without the organization. Natural, accidental, and intentional occurrences that may disrupt conventional operational processes are included in the discussion. The implications of human-machine interaction and how they apply in our wider cyber-security context are also enveloped in the discussion through the aforementioned statement. The protection elements apply to all digital and human assets critical to the information exchange process. Redundancies are also developed to mitigate any adverse effects resulting from irresponsible user behaviour that may jeopardize the integrity of the system. The process operates outwards from the computer network for the entire cyberspace around it.

> ...from events...

The statement recognizes that the computer system and cyberspace are protected from a wide range of characters and events, from human-made to natural, that may jeopardize the integrity of the framework. Defining the range of the circumstances is critical to ensuring that cybersecurity challenges are not only perpetrated by malicious actors as is commonly assumed but may also occur as a result of accidental exposure of sensitive information by users or destruction of the system by natural occurrences.

> ...compromise the integrity and interfere with property rights, resulting in some extent of loss...

The elements of control, ownership, and consequences are highlighted in the statement. The adverse implications of cyber insecurity are outlined to justify the significant investment in guaranteeing the framework's integrity. The statement presents the need for implementing the quality improvement initiative, allowing for a more protracted review and analysis based on specific contextual factors. The term "property rights" is used to encapsulate all types of unauthorized entry and common activities perpetrated by malicious actors, including changing, destroying, or manipulating the functioning of computer systems. Key activities may include exclusion, management, extraction, and removal of any programs, files, or any digital property of the target. Any activity that jeopardizes the continued shamelessness of an organization's operations is regarded as a cybersecurity incident and requires appropriate responses specific to the event and its circumstances. The primary intention of providing a multidisciplinary approach in the definition is to align it with emerging best practices in the field. Cybersecurity is increasingly integrating the input of experts from other fields and appreciating their role in ensuring overall digital integrity.

**Justification Implications and Rationale**

The principal motivation for deriving the new definition is engendering collaborative and interdisciplinary perspectives into cybersecurity. The report endeavours to highlight the need for a broader analysis of the concept and encourage a departure from traditional perspectives, which only focus on system integrity. The paper aims to unify the core elements of cybersecurity and provide a modern outlook in light of the changing behaviour of users and attackers and advances in the sector.

 A more robust approach to ensuring system and network security will be provided with the primary intention of address in traditionally overlooked areas, including malicious-actor motivation and potential disincentivizing strategies. The proposed definition will be substantive if it accomplishes certain core objectives, including:
1. The unification and inclusion of interdisciplinarity in the deconstruction of cybersecurity.
2. The collection and mapping of core concepts defining cybersecurity from existing literature.

The review and analysis of definitions are largely aided by existing literature and an examination of existing best practices. Protracted approach sports adopted in unifying the different elements of cybersecurity and integrating an interdisciplinary perspective to offer a more response. The terms explained in table 2 are mostly drawn from an online repository referred to as the Chambers Thesaurus. The definitions are expected to be as profound as possible to provide a basic understanding of their meaning. The terms have been exploring the cybersecurity context.

| Term | Definition |
| --- | --- |
| **Malicious actor** | - an individual or party interested in achieving a given insidious objective, including gaining unauthorized entry, manipulating, deleting, or destroying a given digital record to disable another party or disrupt its operations. |
| **Collection** | - the process of accumulating or compounding a given resource to achieve a predefined objective. Securing existing infrastructure involves grouping personnel, hardware, and software and allocating functions to address a specific quality improvement opportunity. |
| **Concerting** | - involves combining, arranging, or organizing resources in a coordinated effort to achieve a mutual outcome. Well-defined responsibilities are apportioned to every element, with specific targets being outlined. Consulting in the cybersecurity context involved adding and organizing specific resources, experts, and facilities to ensure network integrity. |
| **Resources** | - an organization's supply of tangible and intangible materials and assets to increase functional efficiency. Resources such as personnel, hardware, and software are critical to effectively responding to threats from unauthorized access to networks and other digital infrastructure. |
| **Personnel** | - human resources required to achieve a given operational objective. Personnel is essential to implementing initiatives as they are responsible for developing automated systems that detect, mitigate, and effectively respond to perceived threats. Personnel is also critical to understanding the extent of loss or damage to property rights, thus, providing a profound and accurate review of expected consequences. |

| **Infrastructure** | - physical facilities and organizational structures in accomplishing a given organizational agenda. Infrastructure is essential to cybersecurity because it provides geographical deterrence and accommodates the necessary hardware and software to sufficiently secure the system. The concept is at the core of ensuring the functional integrity of the digital property of an organization. |
|---|---|
| **Protection** | - the process of identifying, eliminating, resolving, and mitigating the impacts of perceived threats. Protection involves isolating the system from the actions of malicious actors and instituting effective countermeasures to prevent further assaults from the attacker. The concept is the most important attribute of cybersecurity and is the underpinning objective for adopting the construct in conventional cyberspace. |
| **Networks** | - a set of interconnected computer systems intended to achieve an overarching objective. A network consists of electronic devices exchanging data and sharing physical resources required to accomplish organizational tasks. |
| **Integrity** | - the state of being unimpaired or compromised by external or internal influences. Human, software, and hardware elements of a computer network should always exhibit integrity to ensure seamlessness in organizational operations. |
| **Interference** | - disruption of a given entity's conventional activities or operations through the denial of service or introduction of an impediment with the intention of causing loss. Cybersecurity incidents commonly result in interference with normal operations. |
| **Property Rights** | - the legal and theoretical ownership of core tangible and intangible resources by an organization or private individuals. Disruption or manipulation of property rights is one of the core objectives and exercises constituting a cybersecurity incident |

Table 2: terms explained drawn from an online repository referred to as 'The Chambers Thesaurus'.

**Conclusion**

Cybersecurity is often defined as the practice of allocating resources to safeguard the integrity of systems from unauthorized entry or manipulation by malicious actors to ensure minimal disruption of organizational operations. The process involves instituting substantive steps to ensure that any vulnerability exhibited by a computer system or network is sufficiently averted or identified and resolved. Cybersecurity is constituted of several core aspects, including critical infrastructure, applications, networks, the cloud, and the Internet of Things. Critical infrastructure security entails safeguarding unauthorized physical or digital access into protected systems. Punitive legal provisions are outlined on the platform to deter potential attackers and to alert them of consistent monitoring and potential for the activities to be detected and resolved. Physical safeguards may also be integrated, including the isolation of the facilities in remote areas and employing of deterrents such as armed personnel and rigorous access control. Application security is also a major aspect of cybersecurity in evolving end-user software used to manipulate computer systems and networks. The concept presents a unique concern as it exhibits significant vulnerabilities that attackers may exploit. The discussion of cybersecurity also includes network analysis and integrity. A computer system is constituted of several interconnected devices continuously interacting with each other and manipulating data to accomplish a given objective. The terminals present ideal access points for malicious actors who may use them to gain entry to accomplish their objectives. Another emerging practice in contemporary cyberspace is cloud computing. The process entails using off-site storage, database, server, and intelligence capabilities to optimize productivity and gain access to a wide array of third-party software and services. Cloud computing is increasingly adopted in the contemporary digital environment due to conveniences such as the reduced need for physical space and enhanced security. Other elements, such as the internet of things corruption, are often commonly ignored in the contemporary definitions of cybersecurity. The emerging practice entails the interconnection of different devices to increase efficiency and eliminate the need for additional physical infrastructure. Therefore, cybersecurity is interdisciplinary and consists of different aspects and elements involving multiple professional competencies.

The concept has been contested as event-specific and multidisciplinary, thus, warranting a new definition encapsulating the different elements that constitute a security incident. Reinventing the construct is necessary for providing a more uniform interpretation of events, circumstances, and resources critical to understanding it. This report provides a concise, inclusive, and unifying definition of cybersecurity to enhance and enrich research discourse on the topic. Core dielectrics are discussed with the express intention of influencing further analysis by key decision-makers and other special interest groups in the discipline including government, law enforcement, and industry. The approach adopted by scholars, organizations, researchers, and digital security agencies, among others, will be influenced by the broader discussion of the concept provided by this paper. Academia can play a major role in expanding the definition of cybersecurity by ensuring the inclusion of the subjects stated below.

Several salubrious features can be derived from the proposed definition, including:
1. Integrating an element of property rights, loss, and consequence emanating from alienation, exclusion, manipulation, destruction, theft, or unauthorized access to a given system. The perspective is critical as it shifts the discussion beyond traditional interpretations of the concepts such as assets and information security to include a perspective of tangible and intangible value.
2. Including protection is essential to define the wide array of activities encapsulated by cybersecurity. The threats presented by natural, intentional, and unintentional system compromise are also explicitly explained.

Existing literature does not provide a concise and universally profound definition of cybersecurity and its multispectral perspective. The concept is often applied contextually and expanded to include all events and circumstances within a given environment constituting the breach of the integrity of a system. The proposed definition is critical to embracing the multidimensionality of cybersecurity and appreciating its different constituent disciplines. Different elements constituting the contemporary perspectives of systems and network integrity are also included along with the socio-technical dimensions. The author believes that the proposed definition of cybersecurity will be accepted and integrated into contemporary research discourse. Readers will develop an improved understanding of the concept and expand their intellectual Horizons and the openness to collaborate in addressing the increasingly complex threats in modern cyberspace.

# Bibliography

*7 Types of Cyber Security Threats*. (n.d.). (University of North Dakota) Retrieved October 2022, from University of North Dakota: https://onlinedegrees.und.edu/blog/types-of-cyber-security-threats/

Abomhara, M., & Koien, G. M. (2015). Cyber security and the internet of things: vulnerabilities, threats, intruders and attacks. *Journal of Cyber Security and Mobility*, 65-88.

Abouzakhar, N. (2013). Critical infrastructure cybersecurity: a review of recent threats and violations.

Beqiri, E. (2018). The implications of information networking at the pace of business development. *International Journal of Knowledge, 27*, 17-22.

Craigen, D., Diakub-Thibault, N., & Purse, R. (2014). Defining cybersecurity. *Technology Innovation Management Review, 4*(10).

*Cybersecurity vs Network Security vs Information Security*. (2022, April 6). Retrieved from GeeksforGeeks: https://www.geeksforgeeks.org/cybersecurity-vs-network-security-vs-information-security/

Cybersecurity, Critical Infrastructure. (2018). Framework for improving critical infrastructure cybersecurity. *URL: https://nvlpubs. nist. gov/nistpubs/CSWP/NIST. CSWP, 4162018*.

*Education sector is leaving Australians vulnerable to cyber attacks*. (n.d.). Retrieved October 2022, from AISA: https://www.aisa.org.au/public/News_and_Media/Media_Releases/Australians_vulnerable_to_cyber_attacks.aspx

Frankenfield, J., Mansa, J., & Schmitt, K. R. (2022, June 23). *What is Cloud Computing? Pros and Cons of Different Types of Services*. (Investopedia) Retrieved October 2022, from Investopedia: https://www.investopedia.com/terms/c/cloud-computing.asp

Gillis, A. S. (n.d.). *What is the internet of things (IoT)?* (TechTarget) Retrieved October 2022, from TechTarget: https://www.techtarget.com/iotagenda/definition/Internet-of-Things-IoT

*IBM: Consumers Pay the Price as Data Breach Costs Reach All-Time High*. (2022, July 27). Retrieved October 2022, from IBM: https://newsroom.ibm.com/2022-07-27-IBM-Report-Consumers-Pay-the-Price-as-Data-Breach-Costs-Reach-All-Time-High

*Legal consequences of a cyber attack*. (2022, March 9). Retrieved October 2022, from Harper James: https://harperjames.co.uk/article/legal-consequences-of-a-cyber-attack/

Liu, X., Qian, C., Hatcher, W. G., Xu, H., Liao, W., & Yu, W. (2019). Secure internet of things (iot)-based smart-world critical infrastructures: Survey, case study and research opportunities. *IEEE Access, 7*, 79523-79544.

Maglaras, L. A., Kim, K.-H., Janicke, H., Ferrag, M. A., Rallis, S., Fragkou, P., . . . Cruz, T. J. (2018). Cyber security of critical infrastructures. *Ict Express, 4*(1), 42-45.

Mohammed, A., Alkhathami, J., Alsuwat, H., & Alsuwat, E. (2021). Security of Web Applications: Threats, Vulnerabilities, and Protection Methods. *International Journal of Computer Science & Network Security, 21*(8), 167-176.

Petersen, R., Santso, D., Wetzel, K., Smith, M., & Witte, G. (2020). Workforce framework for cybersecurity (NICE framework).

Reddy, G. N., & Reddy, G. J. (2014). A study of cyber security challenges and its emerging trends on latest technologies. *arXiv preprint arXiv:1402.1842*.

Seemma, P. S., Nandhini, S., & Sowmiya, M. (2018). Overview of cyber security. *International Journal of Advanced Research in Computer and Communication Engineering, 7*(11), 125-128.

Shea, S., Gills, A. S., & Clark, C. (n.d.). *What is cybersecurity?* (TechTarget) Retrieved October 2022, from TechTarget: https://www.techtarget.com/searchsecurity/definition/cybersecurity

Srinivasan, J., & Simna, S. (2017). Disaster Recovery, An Element of Cyber Security-A Flick Through. *International Journal of Management (IJM), 8*(4).

Tamrin, S. I., Norman, A. A., & Hamid, S. (2021). Intention to share: the relationship between cybersecurity behaviour and sharing specific content in Facebook.

*What is Cyber Security?* (n.d.). (Kaspersky Lab) Retrieved November 2022, from Kaspersky: https://www.kaspersky.com/resource-center/definitions/what-is-cyber-security

*What is Cybersecurity?* (n.d.). (Gartner, Inc.) Retrieved November 2022, from Gartner: https://www.gartner.com/en/topics/cybersecurity

*What is Cybersecurity?* (n.d.). (Cisco Systems, Inc.) Retrieved October 2022, from CISCO: https://www.cisco.com/c/en/us/products/security/what-is-cybersecurity.html

*What is Network Security?* (n.d.). (Check Point Software technologies) Retrieved October 2022, from Check Point: https://www.checkpoint.com/cyber-hub/network-security/what-is-network-security/

Wong, M. W. (2007). Cyber-trespass and 'unauthorized access' as legal mechanisms of access control: Lessons from the US experience. *International Journal of Law and Information Technology, 15*(1), 90-128.

Yadav, D., Gupta, D., Singh, D., Kumar, D., & Sharma, U. (2018). Vulnerabilities and security of web applications. *2018 4th International Conference on Computing Communication and Automation (ICCCA)* (pp. 1-5). IEEE.